\documentclass[pss]{wiley2sp} 
\usepackage{amsmath}

\begin{document}

\title{Study of leakage currents in PC CVD diamonds as function of the magnetic field}

\titlerunning{Study of leakage currents in PC CVD diamonds}

\author{%
  S. M\"uller\textsuperscript{\Ast,\textsf{\bfseries 1},\textsf{\bfseries 2}},
  W. deBoer \textsuperscript{\textsf{\bfseries 2}}, M. Schneider  \textsuperscript{\textsf{\bfseries 2}}, A. Sabellek \textsuperscript{\textsf{\bfseries 2}}, M. Schmanau  \textsuperscript{\textsf{\bfseries 2}}, C. R\"uhle  \textsuperscript{\textsf{\bfseries 2}}, T. Schneider  \textsuperscript{\textsf{\bfseries 3}}
  R. Hall-Wilton \textsuperscript{\textsf{\bfseries 1},\textsf{\bfseries 4}}
}
\authorrunning{Steffen M\"uller et al.}

\mail{e-mail:
  \textsf{steffen.mueller@cern.ch}, Phone:
  +41 22 76 72078}

\institute{%
  \textsuperscript{1}\,Cern, Geneva, Switzerland\\
  \textsuperscript{2}\,Inst. f. exp. Kernphysik, Universit\"at Karlsruhe (TH), Wolfgang Gaede-Weg 1, Geb. 30.23, 76131 Karlsruhe, Germany \\
  \textsuperscript{4}\,Forschungszentrum Karlsruhe GmbH, Inst. f. Technische Physik (ITP), Hermann-von-Helmholtz-Platz 1, 76344 Eggenstein-Leopoldshafen\\
  \textsuperscript{4}\,University of Wisconsin, Madison, USA}

\received{XXXX, revised XXXX, accepted XXXX} 
\published{XXXX} 


\abstract{%
%
%
%
\abstcol{%
Polycrystalline diamonds produced with the Chemical Vapor Deposition technique (PC CVD-diamonds)  are regularly used as beam loss monitors in accelerators by measuring the ionization of the lost particles.  In the past these beam loss monitors showed sudden increases in the dark leakage current without beam losses and these erratic leakage currents were found to decrease, if magnetic fields were present. Here we report on a systematic }{study of leakage currents inside a magnetic field. The decrease of erratic currents in a magnetic field was confirmed. On the contrary, diamonds without erratic currents showed an increase of the leakage current in a  magnetic field perpendicular to the electric field for fields up to 0.6T, for higher fields it decreases. A preliminary model is introduced to explain the observations.
}
}

%
%

\maketitle   

\section{Introduction}
Modern particle accelerators, such as the Large Hadron Collider (LHC \cite{bib1}) at CERN can store unprecedented amount of energy in their circulating beams. Even the loss of a small fraction of this energy can cause serious damage to the accelerator facilities or the installed experiments such as the Compact Muon Solenoid (CMS \cite{bib2}). This means, that a sophisticated and gap-less beam monitoring is indispensable in order to safely operate such a machine. The LHC and CMS will be introduced in more detail below. 

The LHC monitoring system is called Beam Loss Monitoring system (BLM \cite{bib3}) and consists of approximately 4000 ionization chambers, which are mounted onto the quadrupole magnets. These chambers are about 60cm long, so that it is not possible to mount them inside the experimental areas, such as the CMS cavern. In order to provide a gap-less monitoring, PC CVD-diamond detectors were used instead of the ionization chambers inside the experiments. Diamond detectors have the same working principle as ionization chambers, allowing use of the same readout electronics. There are also some other advantages compared to ionization chambers like compact size, fast signal response and low bias voltage. In addition they are relatively inert to environmental conditions. Some PC CVD sensors tend to show erratic currents, i.e. a spontaneous increase in leakage current, typically from the pA range to the nA range. After a while a steady state is reached and the current remains at a high level until the bias voltage is turned off. Presumably these currents are caused by an internal breakdown of the electric field within the crystal. More details of erratic dark currents can be found in  \cite{bib9}. In all reported cases the erratic current practically disappeared in magnetic fields of the order of a few tenths of a Tesla  \cite{bib8}.

To understand the influence of the orientation and the magnitude of the magnetic field on the leakage current a PC CVD diamond sensor was put in a laboratory magnet with a field up to 10 T. This spare diamond from the CMS beam monitor (BCM2) had not exhibited erratic currents.  Surprisingly, the leakage current in this diamond increased with increasing magnetic field perpendicular to the electric field contrary to the erratic currents. With the magnetic field parallel to the electric field only small effects were seen. A preliminary model based on the grain boundaries in the polycrystalline diamond sensors is presented in the last section. 

\paragraph{The Large Hadron Collider}
The LHC is a proton-proton collider located in Geneva - Switzerland. With a circumference of 27km and a magnetic dipole field of 8T, a maximum particle energy of 7TeV can be reached. 2808 particle packets with 1.15$\times 10^{11}$ protons will be injected with an energy of 0.45TeV per beam. At the nominal energy the total stored energy in both beams is around 700MJ. 

\begin{figure}[h]%
\begin{center}
\includegraphics*[width=0.8\linewidth,angle=-0]{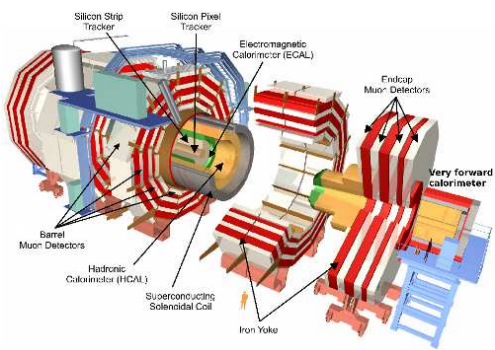}
\end{center}\caption{%
Exploded view of the CMS detector. The positions of the various sub-detectors are indicated.}
\label{cms}
\end{figure}

\paragraph{Compact Muon Solenoid - CMS}
CMS is one of the four main experiments at the LHC and a multipurpose detector for the physics at the TeV scale with a solenoidal magnetic field of 3.8T. Several sub-detectors like the silicon tracker, calorimeters and muon chambers are arranged concentrically around the beam pipe, thus allowing to measure all relevant parameters of the produced particles by the proton-proton collisions. These detectors are shown schematically in fig. \ref{cms}. CMS has a diameter of about 15 meters, a length of 21.5m and a weight of more than 12 thousand tons.

\paragraph{Magnetic field in CMS}
The nominal magnetic field of 3.8 Tesla in CMS is generated by a superconducting coil with a diameter of 6m and a length of 12.5 m. After the magnet was turned on, the magnetic field map was measured, showing that a significant field exists in the forward region. The BCM2 diamonds are exposed to a magnetic field of up to 0.7T.  

\section{Beam and Radiation Monitoring for CMS (CMS-BRM)} 
The colliding beams within CMS are continuously monitored in order to detect the onset of adverse beam conditions before any of the CMS sub-detectors are endangered. A modular design was chosen for the BRM system \cite{bib12}, to have a very flexible and versatile system. Up to seven different subsystems are monitoring specific parameters of the beam environment. While some of them are only for (long-term) monitoring, others can send a signal to the LHC to dump the beam, whenever thresholds are exceeded. In the following, one of the beam condition monitors of CMS, the BCM2 system \cite{bib7}, will be introduced as such a system. 

\subsection{Beam Conditions Monitor 2 (BCM2)}
\paragraph{System overview}
In total 24 PC CVD diamonds are positioned at $Z=\pm 14.4$m from the interaction point (IP) in two radii around the beam pipe. A DC coupled current meter measures the particle induced signal current which are compared each 40 $\mu$s with a threshold value. Whenever this value is exceeded a beam abort signal is send to the LHC. The diamonds are mounted inside four Aluminum structures and fixed onto a support table in the hadron forward calorimeter.

%

\paragraph{Diamond detector package}
The diamond detectors used by BCM2 are 10x10mm$^2$ polycrystalline CVD-diamond with a thickness of 400 $\mu$m. They are standard detector grade sensors from E6\footnote{E6-web address: http://www.e6.com}. For detector grade diamonds a part of the substrate side is removed in a lapping process, to ensure to have relatively large grains remaining. A 0.1 $\mu$m thick Tungsten-Titanium metallization with an active area of 9x9 mm$^2$ was done at Rutgers University. The collected charge is usually significantly smaller than the expected charge for a given thickness of the detector, because of recombination of the charge carriers, mainly at the grain boundaries of the polycrystalline samples. The charge reduction is usually expressed by a reduced (effective) thickness of the detector, usually called the charge collection distance (CCD), which  was measured with Alpha-spectroscopy to be 234 $\mu$m  for an applied field strength of 1 V/$\mu$m, i.e. only about 60\% of the generated charge is collected. Before measuring the CCD the detectors were irradiated until they reached a stable CCD. This increases the CCD somewhat, because the traps, where recombination of the charge carriers can occur, become partially filled. 

\begin{figure}[h]%
\begin{center}
\includegraphics*[width=0.5\linewidth,angle=-0]{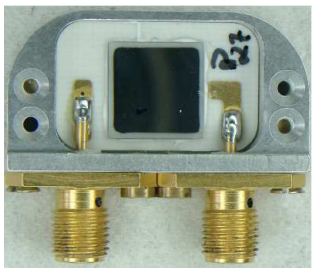}
\end{center}
\caption{%
Close up view of a mounted BCM2 diamond. All diamonds were glued and wire bonded onto a glass-ceramic baseplate. SMA connectors provide a reliable connection to the readout electronics.}
\label{P27}
\end{figure}

A fully covered and compact Aluminum housing provides good shielding against noise pick-up. Inside the housing, the diamond is held with thermoplastic glue \cite{bib11} on a radiation hard glass-fiber enhanced ceramic baseplate \cite{bib14}. Wire bonds are connecting the bias voltage and signal cables to the diamond electrodes. A photograph of an opened detector-package can be seen in figure \ref{P27}. After mounting a long-term test of several days to a week is performed to see whether the diamond shows a tendency to erratic leakage currents or not.

\begin{figure}[h]%
\includegraphics*[width=\linewidth]{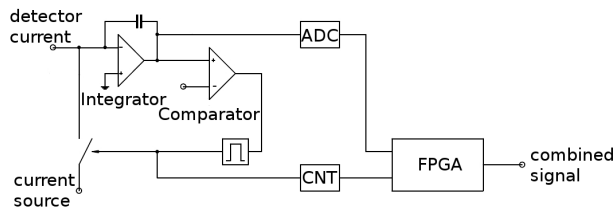}
\caption{
The simplified schematics of the BLM-readout electronics for one channel. See text for details.}
\label{fig:integrator}
\end{figure}

\paragraph{Readout electronics}
The diamond detectors are read out with a current to frequency converter, as used for the LHC-ionization chambers \cite{bib4}. This technique allows a high dynamic measurement range between 2.5pA and 1mA (8 decades -- 160dB). A simplified schematic of one readout channel can be seen in fig. \ref{fig:integrator}. The detector current discharges the capacitor in the integrator circuit. Whenever the voltage on the capacitor drops below  a threshold voltage, a comparator will trigger a current source, which will charge the capacitor with a well defined charge. The number of recharges is counted, thus providing a measure of the detector current. During the recharge the detector current continues to discharge the capacitor, resulting in a dead-time free measurement. The counter measures the discharge in discrete steps. The precision is improved by
an analogue to digital converter (ADC), which samples the integrator voltage during the discharge.  The recharge counts and the ADC values are merged with the on-board processor (FPGA) into one signal value: the most significant bits are given by the counts, the least significant bits by the fractions of a count determined by the ADC. The merged value is transmitted every 40 $\mu$s to a higher level readout board called DAB64x via optical fibers \cite{bib5} \cite{bib6}.

\section{Effects of magnetic field in BCM2 data}

\begin{figure*}[h]%
\subfloat[Onset of erratic dark current in P31 measured over a few days. Current remained stable at 5nA, although this is way higher than a normal leakage current, this still doesn't endanger normal operation of BCM2.]{\label{p31leak}%
\includegraphics*[height=.32\textwidth,angle=-90]{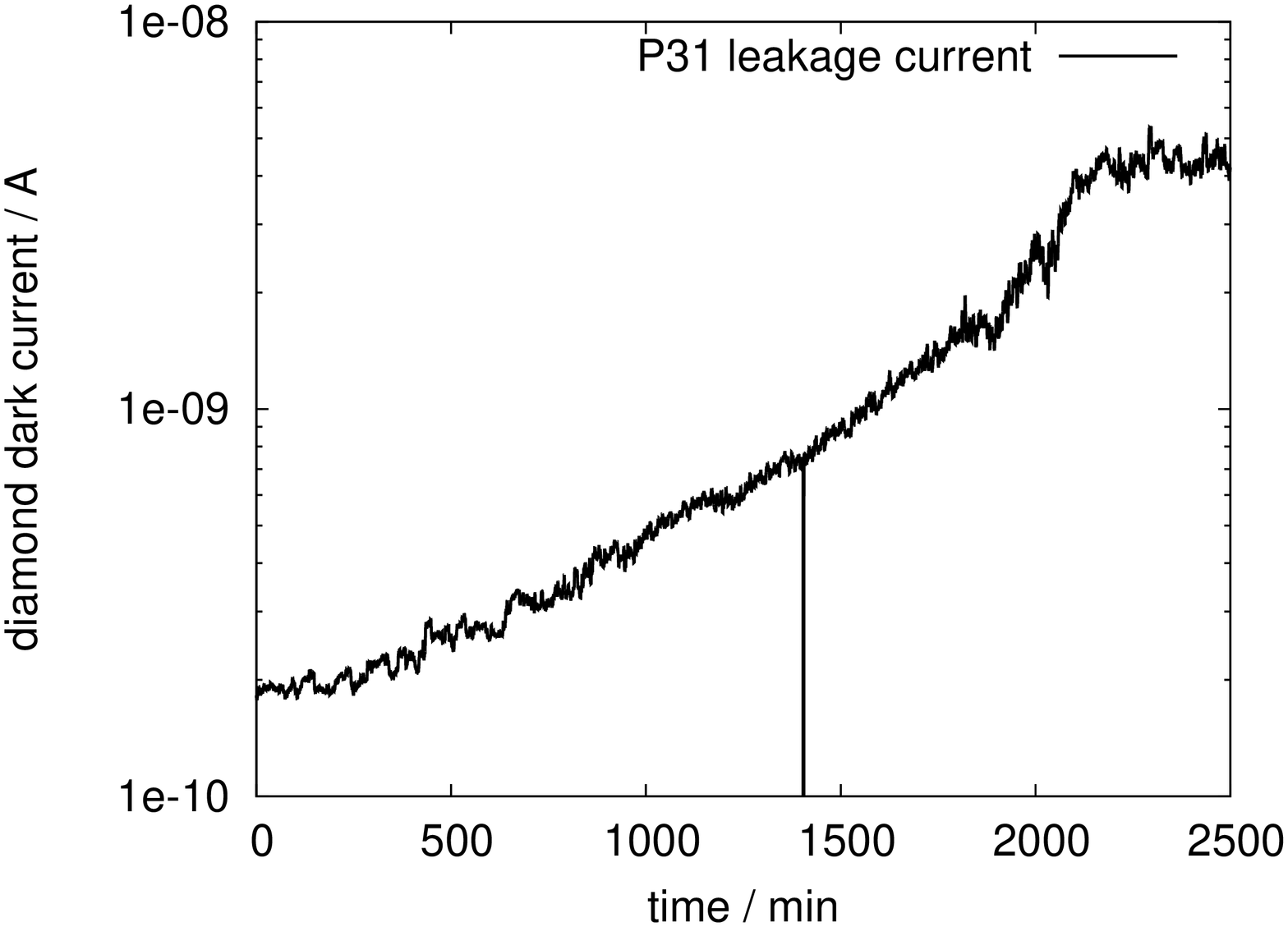}}\hfill
\subfloat[Suppression of erratic leakage current during ramp-up of the magnetic field. The current dropped from 5nA down to 20pA. Shown is the nominal solenoid field strength, the actual field strength and direction at the diamond location is unknown.]{\label{p31ru}%
\includegraphics*[height=.32\textwidth,angle=-90]{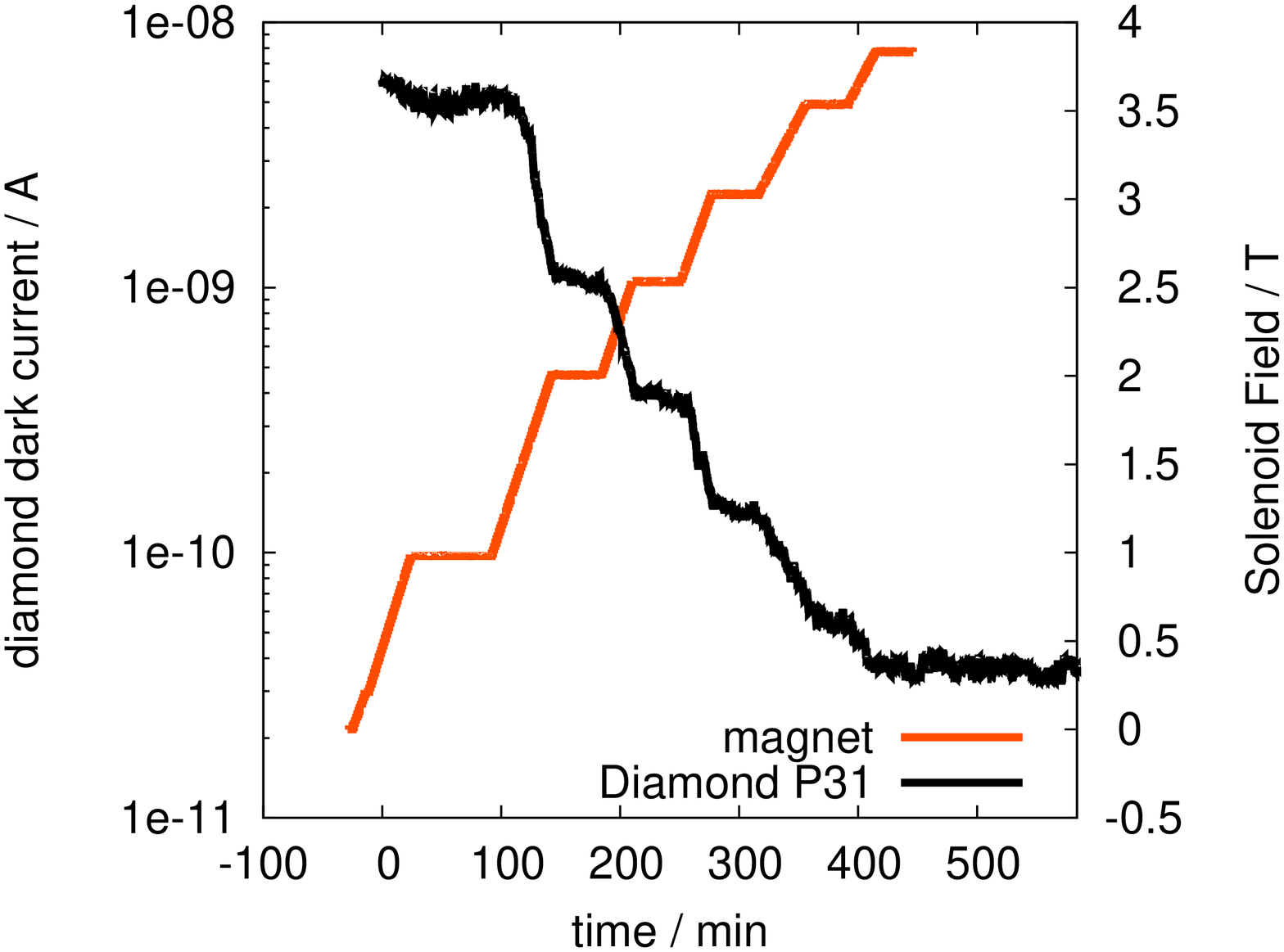}}\hfill
\subfloat[Onset of erratic leakage current during ramp down the magnetic field. The behavior of P31 is reversible, i.e.  current reaches the same current as before after ca. 13h. Before the field was switched off, there was a stable field for about five days, the leakage current showed no variations during that time.]{\label{p31rd}%
\includegraphics*[height=.32\textwidth,angle=-90]{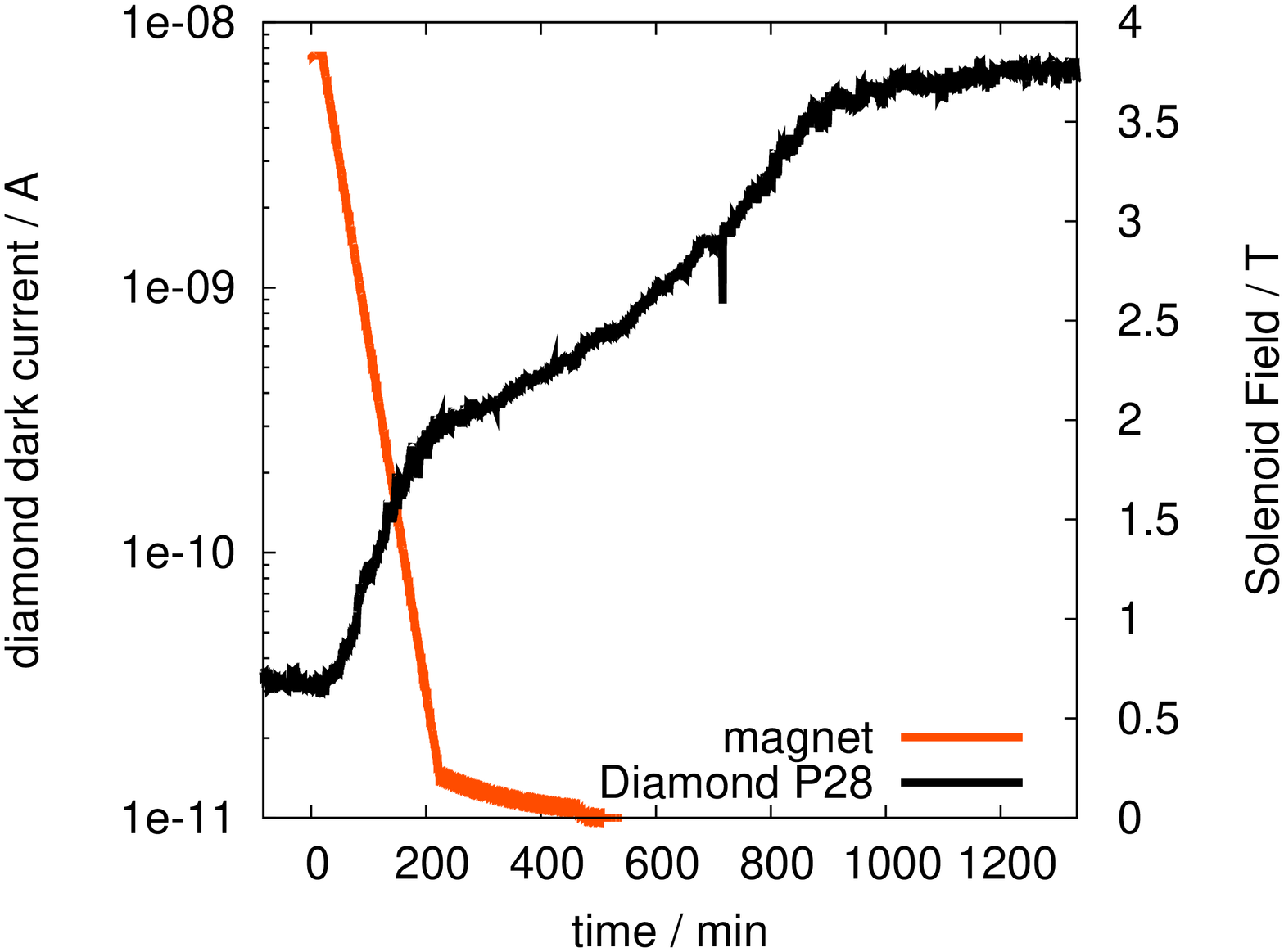}}%
\caption{%
Leakage current of diamond P31 during change of magnetic field. This diamond developed an erratic dark current before the magnetic field was turned on.}
\label{p31}
\end{figure*}

\begin{figure*}[h]%
\subfloat[Leakage current of P28, shown is data for one day. The diamond is very stable over time, no significant variations seen.]{\label{p28leak}%
\includegraphics*[height=.32\textwidth,angle=-90]{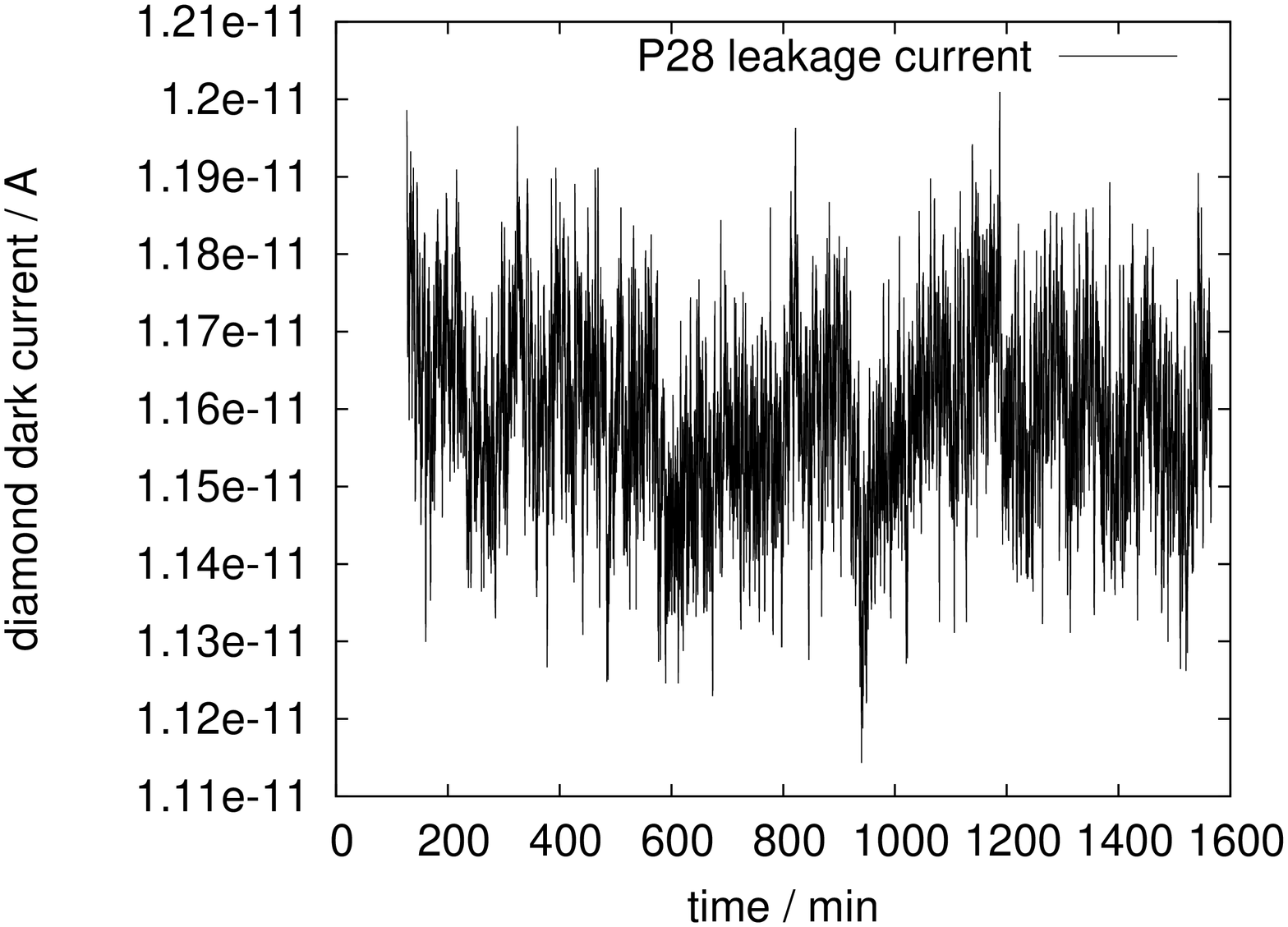}}\hfill
\subfloat[Change in leakage current during ramp up of the  CMS solenoid. The current suddenly increased by 1pA.]{\label{p28ru}%
\includegraphics*[height=.32\textwidth,angle=-90]{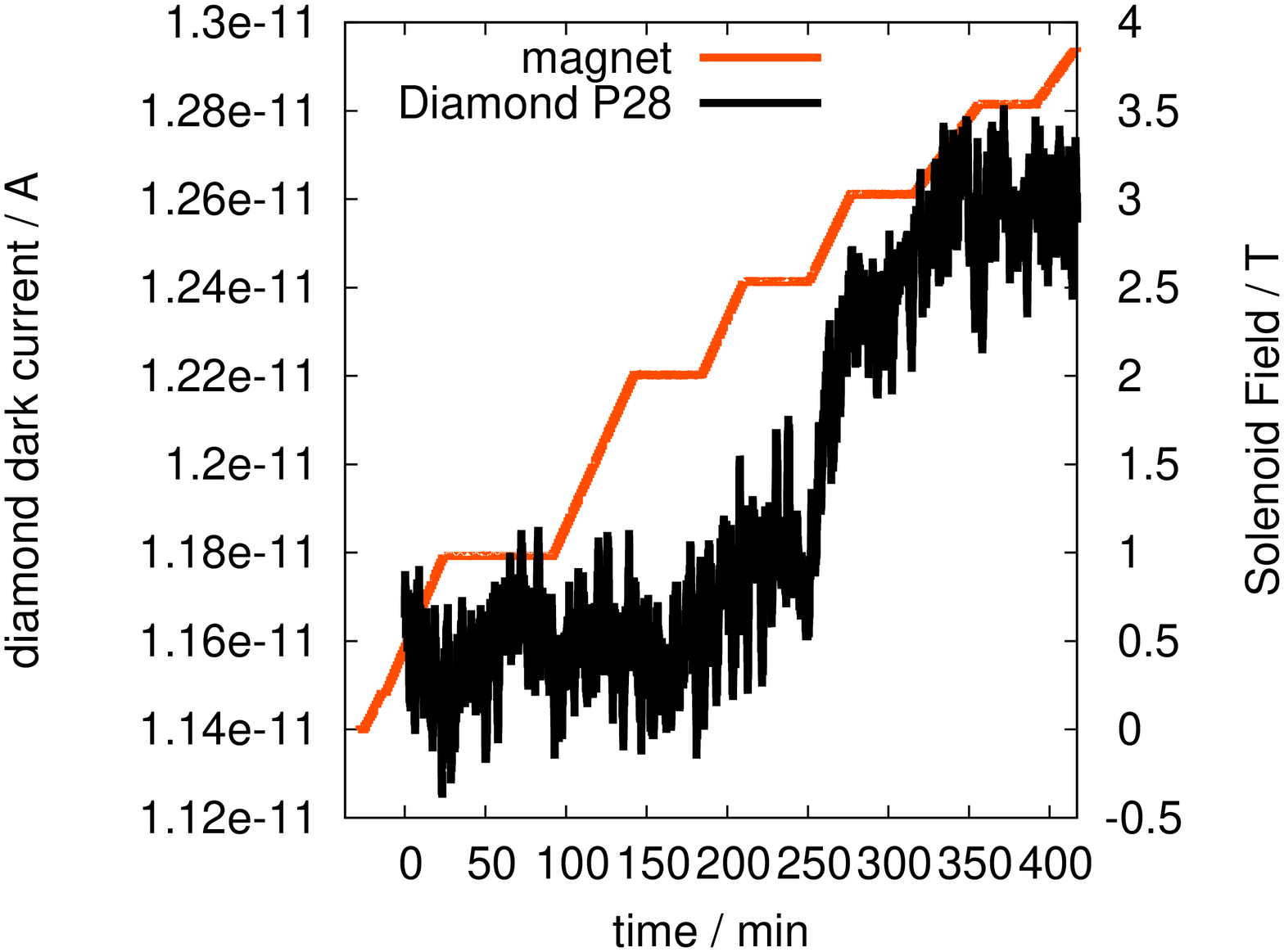}}\hfill
\subfloat[Change in leakage current after ramp down of B-field. The current decreases again by 1pA at the same magnetic field, where it increased before.]{\label{p28rd}%
\includegraphics*[height=.32\textwidth,angle=-90]{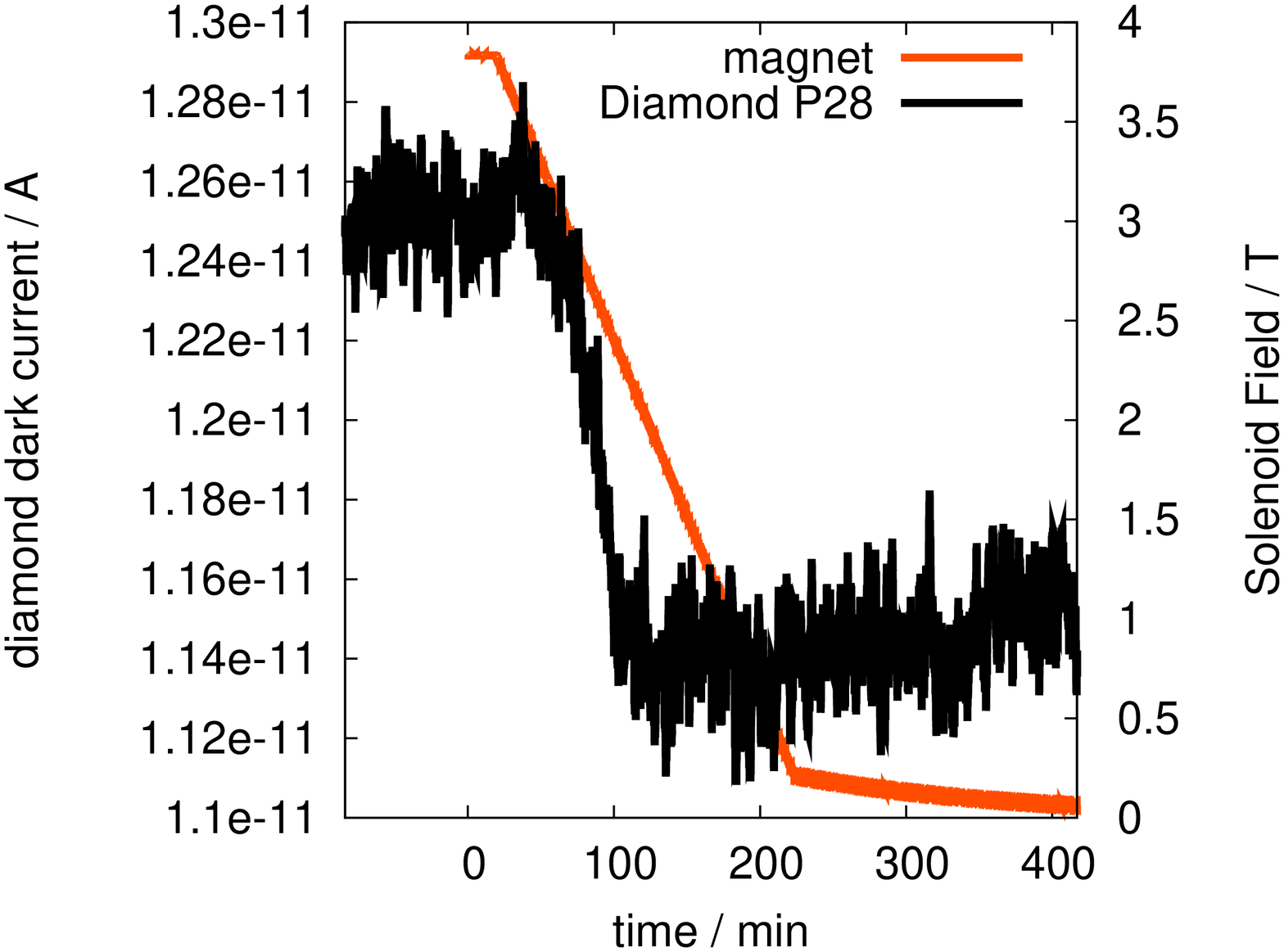}}%
\caption{%
Leakage current of diamond P28 during change of magnetic field. This is a  diamond without erratic currents. }
\label{p28}
\end{figure*}

The diamonds were carefully selected and quality assured. Only diamonds showing no erratic currents during  long term tests of several weeks were used for BCM2. Neverthelss, after a few weeks of operations   one of the 24 diamonds  developed an erratic leakage current. No attempt was made to lower the bias voltage in order to see how it will develop over time and in a magnetic field.

During the CRAFT\footnote{Cosmic Rays At Four Tesla} run starting from 16th October 2008 two different effects were seen in the leakage current of a few diamond detectors. The actual field strength and direction at the diamonds position is not known, but believed not to be stronger than 0.7T. Given the amount of iron around the diamond detectors it is also  hard to estimate the field direction.
 
The characteristics of the diamonds mentioned in the following paragraphs are shown in table \ref{diapar}. While the CCD is  similar in these samples, the measured leakage current for P27 shows a strong dependence on the polarity of the electric field. This is also the case for P28. The processing and metallization of the diamonds  were done identically for all samples. Therefore the asymmetry for some characteristics is presumably caused by the crystal itself. The diamonds were already polished on both sides, so it was not possible to identify which side is the growth or substrate side.  

\begin{table}[b]
  \caption{Characteristics of diamonds used here}
  \begin{tabular}[htbp]{@{}r|lll@{}}
    \hline
	Diamond & CCD at 1$\frac{V}{\mu m}$ &current at 0.5$\frac{V}{\mu m}$&current at -0.5$\frac{V}{\mu m}$\\
\hline
	P27&235$\mu$m&230pA&10pA\\
	P28&237$\mu$m&9pA&53pA\\
	P31&225$\mu$m&17pA&29pA\\
    \hline
  \end{tabular}
  \label{diapar}
\end{table}

\subsection{P31 - a diamond with erratic dark current}
Diamond P31 was the only diamond which developed an erratic dark current. The onset of this erratic current is shown in fig. \ref{p31leak}. The current rises over several days to 5nA, where it remained stable. Whereas 5nA is a relatively high current for a diamond detector, it does not endanger the safe operation of BCM2, as the abort threshold is several orders of magnitude higher.

During the ramp up of the magnet, the erratic dark current dropped from 5nA in a correlated way down to 20 pA, shown in fig. \ref{p31ru}. The magnet remained at the nominal field of 3.8 T for about 5 days. During that time no significant variations in the leakage currents were seen. Whilst the magnet was ramped down again, the leakage current rose as  shown in fig. \ref{p31rd}. One can see that the leakage current reaches the same value as it had before the magnet was turned on, but only after 12 hours. There is still a correlation in the leakage current behavior, but delayed with a long term time constant. 

\subsection{P28 - a well behaved diamond}
Apart from P31, some other diamonds were showing very marginal, but still measurable variations in leakage current of about 1pA. As example we show the behavior of P28, which had a leakage current without significant variations before the magnet operation. This is shown in fig. \ref{p28leak} for a time period of one day.
During the magnet ramp up, the current rose by 1pA from about 11.6pA to 12.6pA and remained stable till the magnet was ramped down again, which is shown in fig. \ref{p28ru}. The correlation in the rise with the magnetic field is not as strong as with P31, but still visible. The leakage current of P28 went back to the previous value of about 11.6pA whilst the magnetic field was switched off. It seems to be a completely reversible effect with no or only a very small time constant.

\section{Lab Measurements at the Jumbo facility}
Although both observations of the leakage current behavior of diamonds in the magnetic field of CMS do not endanger the functionality of BCM2, additional measurements have been done to further understanding of the diamond response at low current. These were done with a spare diamond at the Jumbo magnet test facility at ITP, Forschungszentrum Karlsruhe. The Jumbo facility provides a magnetic field of up to 10 T in a 10 cm bore. The device under test can be cooled with cold N$_2$ gas.

\paragraph{Measurement procedure}

The diamond P27 was put into the field in two directions: the first measurement was done with the B-field perpendicular to the electrical field E, while for the second measurement the diamond was rotated by 90 degrees so that the B-field was parallel to the electrical field.  For both directions the magnetic field was increased in steps; after each step the magnetic field was kept constant for about three minutes

 The leakage current was measured with the current to frequency converter discussed before. The unused readout channels showed no effects from the magnetic field, so that the measured effects have to be caused by the diamond.  The measurements were done with an applied bias voltage of 400 V, i.e. 1 V/$\mu$m. One additional test was done with 700 V.

\begin{figure}[h]%
\begin{center}
\includegraphics*[height=0.9\linewidth,angle=-90]{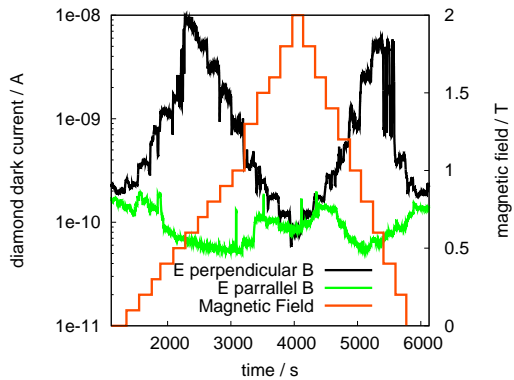}
\end{center}
\caption{%
Raw data of leakage current measurements of P27 shown for E perpendicular B and E parallel B.}
\label{rawdata}
\end{figure}

\paragraph{Results}
The raw data are shown in figure \ref{rawdata}. In both cases large changes in leakage current were observed. The leakage current for E perpendicular B increased to a maximum at a B-field of around 0.6T. At higher fields the current decreases. For E parallel B one can see a smaller and qualitatively different effect: the current first drops with increasing B-field and returns to the initial value at higher fields. The leakage current   shows a similar behavior, wenn the field is reduced, so the effects are reproducable. In the following section the order of magnitude increase in the leakage current for E perpendicular B will be confronted with a model. However, the small but reproducable effect for E parallel B will not be discussed anymore, since such small effects are likely to be sample dependend. The large effect for E perpendicular B could actually be reproduced in a second detector after completion of this paper. A measurement with a singe crystalline CVD (SC CVD) sample did not show any significant effect. This is not expected in the model discussed below.

\begin{figure}[h]%
\begin{center}
\includegraphics*[height=0.9\linewidth,angle=-90]{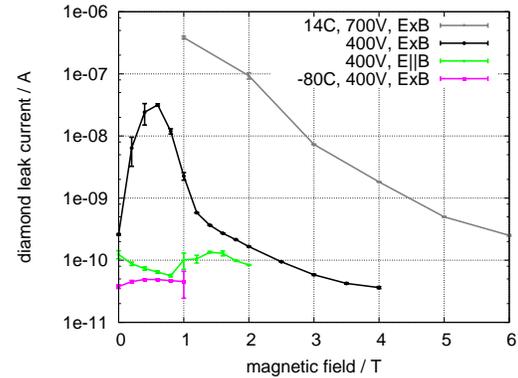}
\end{center}
\caption{%
Results of the leakage current measurements for various B-field directions and E-field strengths. Shown is also a measurement at a temperature of -80C.}
\label{ibcomp}
\end{figure}

Figure \ref{ibcomp} shows all measured leakage current data as function of the magnetic field. At lower temperatures the effect is suppressed and the maximum of the leakage current is reached at a lower B-field. At high E-field the leakage current is higher as expected, but shows the same qualitative behavior for high B-fields.

\section{Preliminary Model}
In the following paragraphs a possible model is presented, which could explain the measured effects. The bandgap of diamond is too large to thermally generate a leakage current, therefore the charge carriers causing the leakage current are believed to be injected from the metal electrodes into the diamond. Electron injection from a metal contact into the conduction band of diamond is expected for a  sufficiently high concentration of defects, in which case the electrons can tunnel through the width of the potential barrier of the Schottky contact \cite{bib9a}. For PC CVD diamonds a high concentration of impurities and / or defects is expected only at the growth side, so the injection of electrons is expected to happen only for one polarity, in agreement with observation. For single crystals the defect concentration is too small in which case the barrier is too broad, so the observed effects should not happen, again in agreement with observation.

Surface effects are unlikely, both because guard rings did not suppress current in \cite{bib9}, and two independent diamond samples showed the effects of figure \ref{ibcomp} independently. The drift parameters of electrons in diamond are shown in table \ref{driftpara}, they were obtained using the parameters from \cite{bib10a,bib10}. Given the large uncertainty in these samples, we ignore the difference of the Hall- and drift-mobility. Details about Lorentz angles in particle detectors can be found in Ref. \cite{bib10b}.

\begin{table}[b]
  \caption{Drift parameters for electrons in PC CVD diamond.}
  \begin{tabular}[htbp]{@{}rl@{}}
    \hline
	Drift mobility& $\mu= 100 - 1000 \frac{cm^2}{VS}$ \\
	Drift velocity& $v_D=\mu E = 1000 \frac{cm^2}{VS} 1 \frac{V}{\mu m}=10^7 \frac{cm}{S}$\\
	Cycl. frequency& $\omega = \frac{eB}{m}=1.7 \times10^{11} \frac{rad}{Ts}$ \\ 
	Radius at 1T& $r=\frac{mv}{eB}=10^7 \frac{cm}{S}\times1.7 \times 10^{11} \frac{rad}{Ts}=1\mu m$\\
	Lorentz-angle&$\phi=\arctan(\mu B) = 0.6^\circ - 6^\circ$ for one Tesla\\ 
	Diffusion length&$l=1.6\mu m$\cite{bib10a}\\
    \hline
  \end{tabular}
  \label{driftpara}
\end{table}

\begin{figure}[t]%
\subfloat[B=0]{\label{b0}%
\includegraphics*[width=.155\textwidth,height=1.6cm]{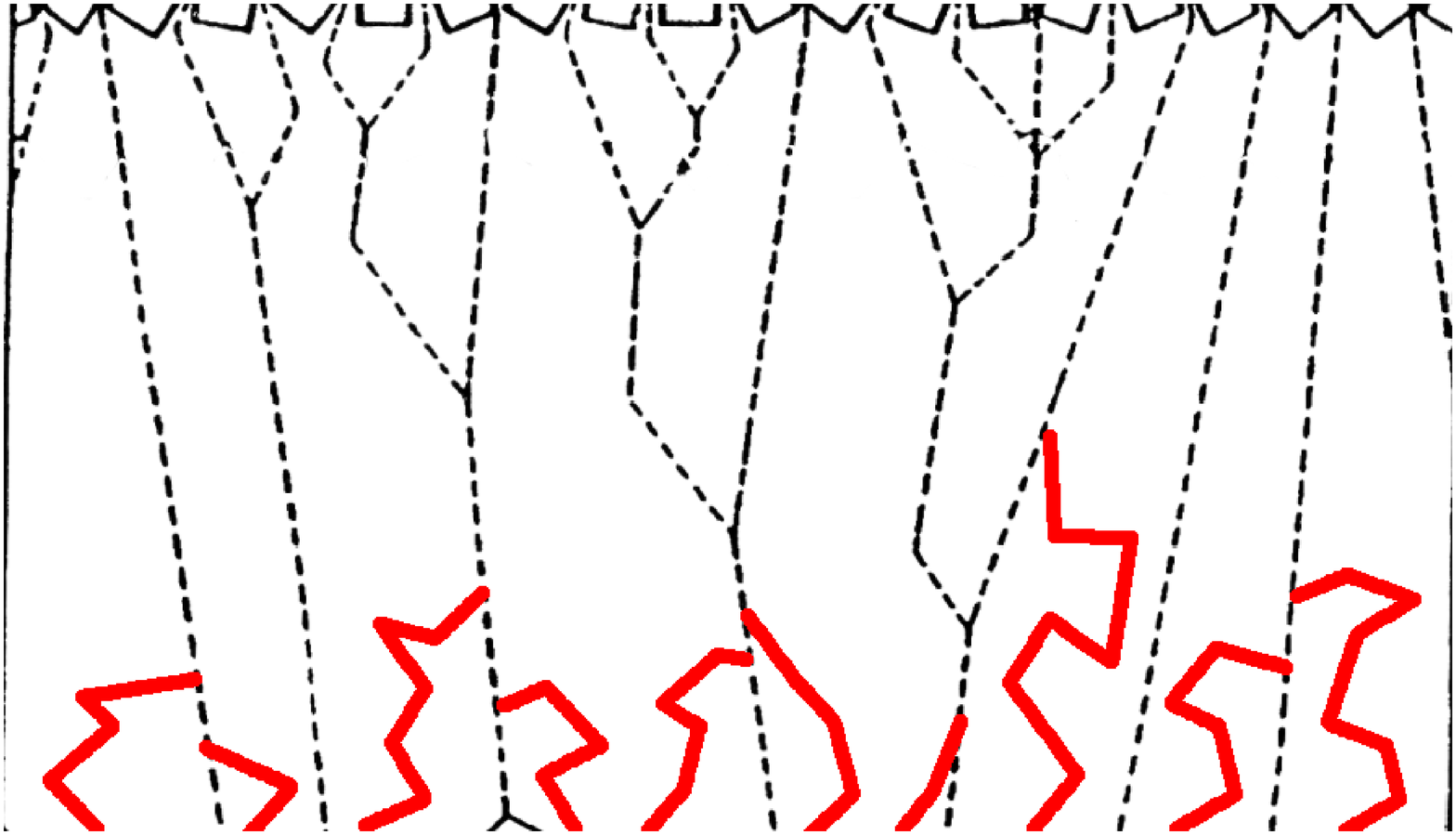}}\hfill
\subfloat[B$\approx$ 1T]{\label{b1}%
\includegraphics*[width=.155\textwidth,height=1.6cm]{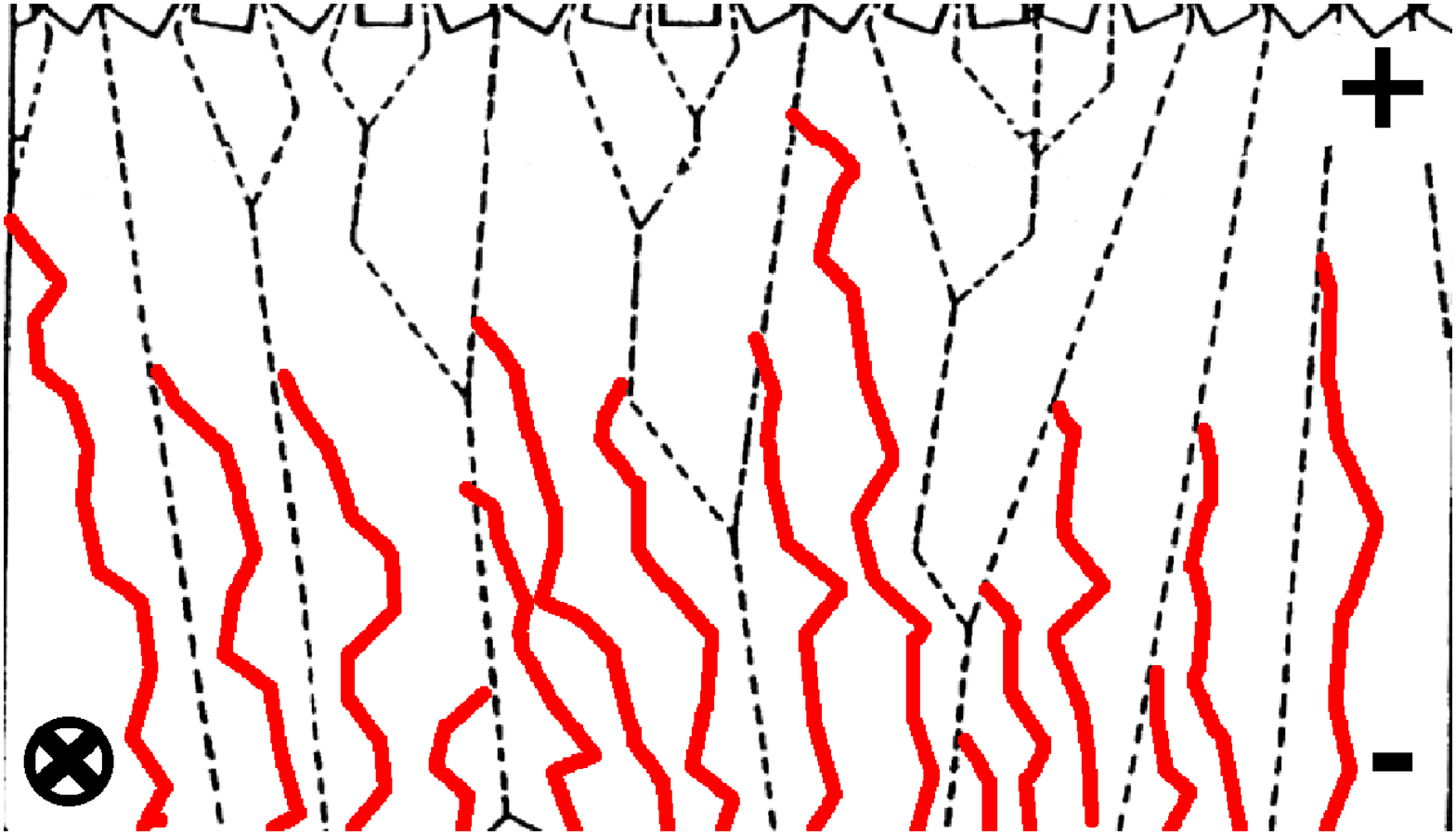}}\hfill
\subfloat[B $>$ 1T]{\label{bhigh}%
\includegraphics*[width=.155\textwidth,height=1.6cm]{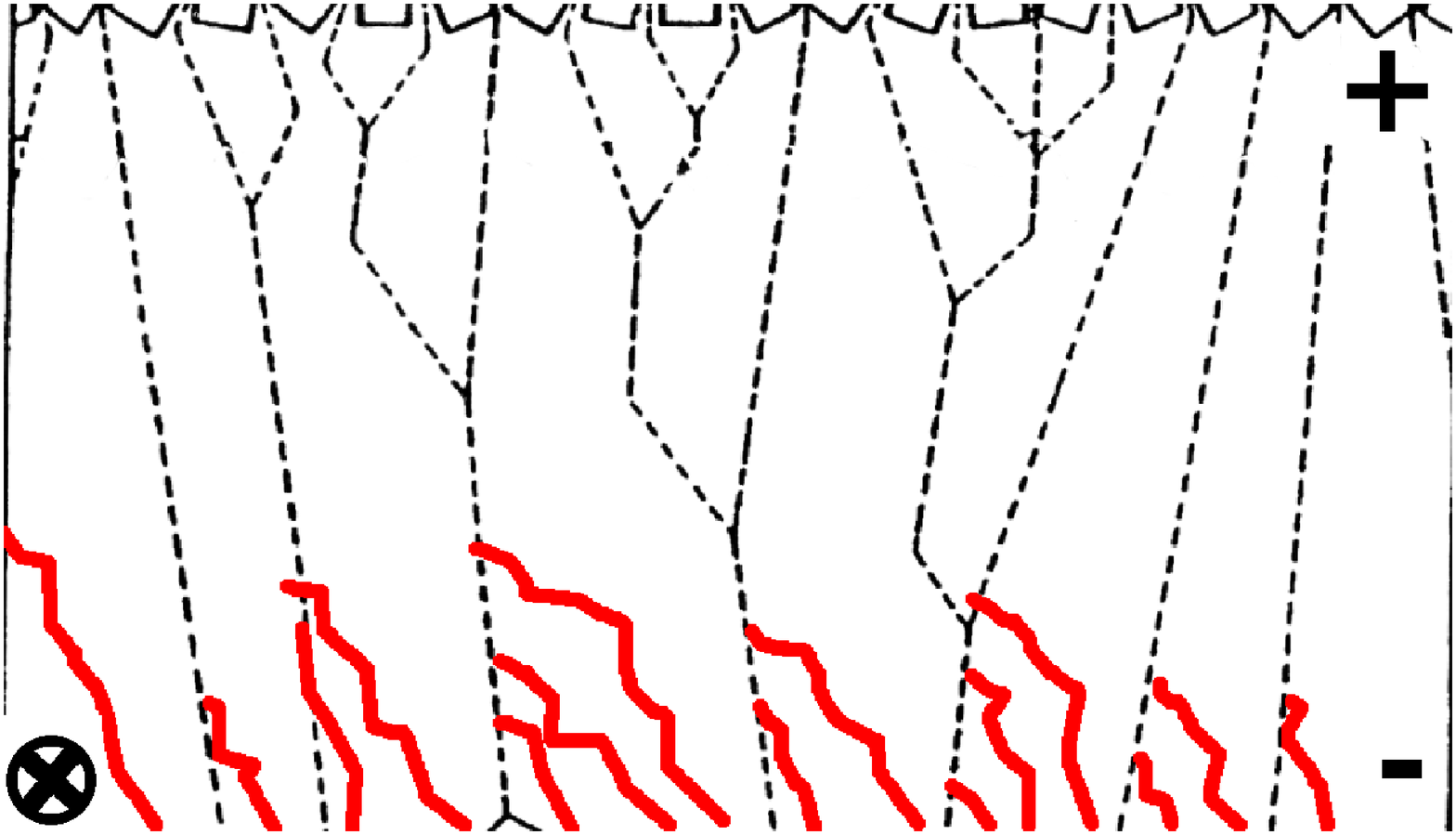}}\hfill%
\caption{%
Sketches of the length of the electron drift (solid lines) in the indicated B-field. The dotted lines indicate grain boundaries of PC CVD diamonds.}
\label{threesubfigures}
\end{figure}

At zero magnetic field the electrons are drifting along the electric field as shown in fig. \ref{b0}. In average they scatter isotropically every 1.6$\mu$m with the Carbon atoms. Therefore there is a high probability that the electrons hit a grain boundary, where they can recombine and therefore no longer contribute to the leakage current. Note that the grain boundaries have typical distances of a few $\mu$m at the nucleation side. At approximately one Tesla, the transversal drift is suppressed by the magnetic field, which forces movement of charge carriers on circles with a cyclotron radius of about 1 $\mu$m.   Therefore the chances of recombining at a grain boundary is reduced and the electrons can drift longer, thus increasing the leakage current  correspondingly. The increased drift paths have been sketched in fig. \ref{b1}. There is also a small Lorentz angle at 1 T, but given the diameter of the grains of typically a few $\mu$m to a few tens of $\mu$m \cite{bib10},  a small angle does not have a big impact. For higher B-fields the Lorentz angle starts to become large, so  the charge carriers can only drift a small distance before hitting a grain boundary, as sketched in fig. \ref{bhigh}. 

The increase in current can be significant if the length of the individual crystals a PC CVD diamond is much larger than their diameter, which is the case. At lower temperatures the Hall-mobility and with it the Lorentz angle increases. Therefore the effects described above already occur at lower magnetic fields, as shown in figure \ref{ibcomp}.

\section{Conclusion}
During the magnet test of CMS two different effects for the leakage currents of the BCM2 diamond detectors could be seen. The leakage current of one diamond showing an increased erratic dark current, which  decreased in a magnetic field. This was expected and already seen by other experiments like CDF and BaBar. Some other diamonds were showing an increase of leakage current in a magnetic field. This effect has never been seen and was studied under laboratory conditions for the first time. For E {\it perpendicular} B the leakage current increased by an order of magnitude almost linearly for a field up to 0.6 T.  At higher fields the current started to decrease. On the contrary, for E {\it parallel} B the current dropped roughly by a factor two  up to a field of 0.8T from where on it reached the same value again. Two independent detectors showed this behaviour in a very similar way.

A preliminary model based on this data was developed, where grain boundaries combined with the drift of electrons inside a magnetic field are the reason for the measured behavior. The electrons are likely to be injected from the metal contacts, where the high concentration of impurities and/or defects at the nucleation side can reduce the thickness of the barrier between the metal and the conduction band of diamond sufficiently to allow for tunneling of the electrons across this barrier. In this case the effect is expected to be important only for one polarity of the bias voltage in PC CVD diamonds and should be absent in SC CVD diamonds due to the lack of a high density of defects. Both predictions are born out by the data.  It is still unknown whether the signal current of ionizing particles is also influenced by this effect. It would be also interesting to see if this effect could be used to determine the quality of the diamond samples. Finally it should be mentioned that none of the measured effects are believed to endanger the safe operation of BCM2. 

\begin{acknowledgement}
We gratefully acknowledge the support of the BCM2 group for the use of the spare diamond of the BCM2 beam monitoring system, which was metallized by the group of R. Stone at the University of Rutgers. Furthermore illuminating discussions with Christoph Nebel from the Fraunhofer Institute in Freiburg about the preliminary model have been very helpful. 
\end{acknowledgement}

\end{document}